\documentstyle[12pt]{article}
\newcommand{\beq}{\begin{equation}}
\newcommand{\eeq}{\end{equation}}
\newcommand{\bea}{\begin{eqnarray}}
\newcommand{\eea}{\end{eqnarray}}
\newcommand{\beann}{\begin{eqnarray*}}
\newcommand{\eeann}{\end{eqnarray*}}
\newcommand{\pder}[2]{\frac{\partial #1}{\partial #2}}

\begin{document}
\title{Elastic forces that do no work and the dynamics of fast cracks}
\author{Fernando Lund \\
Departamento de F\'\i sica \\ Facultad de Ciencias F\'\i sicas y
Matem\'aticas \\ Universidad de Chile \\ Casilla 487-3, Santiago, Chile}
\date{}
\maketitle
\begin{abstract}
Elastic singularities such as crack tips, when in motion
through a medium that is itself vibrating, are subject to forces orthogonal to
the direction of motion and thus impossible to determine by energy
considerations alone. This fact is used to propose a universal scenario, in
which three
dimensionality is essential, for the dynamic
instability of fast cracks in thin brittle materials.
\end{abstract}

The dynamics of crack tips is of particular relevance to the fracture of
materials\cite{freund} and to earthquake rupture\cite{qk}. In addition,
recent improvements in instrumentation techniques have revealed unstable crack
dynamic behaviour reminiscent of what is observed in other physical systems
driven far from equilibrium\cite{p1,p2,p3,p4}. These experiments have taken a
close quantitative look at the crack-tip velocity fluctuations, ultrasound
emission and surface heterogeneities in glass and plexiglass associated with
long standing puzzles in the tensile cracking of very brittle solids\cite{old}.

The theoretical effort directed at an understanding of these phenomena has been
largely based on two dimensional modelling. It includes analysis based on a
continuum approach that considers a cohesive region at the crack
tip\cite{langeretal} and numerical computations using lattice\cite{marderetal}
as well as molecular dynamics\cite{abrahametal} techniques. However,
ultrasound\cite{p2,p3}, surface topography\cite{p1,p3} and velocity
fluctuation\cite{p1,p3} measurements clearly involve length scales comparable
or
smaller than sample thickness, strongly suggesting that three dimensionality
cannot be ignored. The possible relevance of results obtained in a study of
the three dimensional propagation of a crack within an heterogeneous model
elastic solid have been recently pointed out\cite{riceetal}.

The purpose of this paper is to point out that, when the motion of an elastic
singularity deviates from a straight line, energy balance considerations are
insufficient to determine the singularitiy's motion
since (in two dimensions) they provide only one equation to
determine the dynamics of two degrees of freedom. Thus, in addition, the
equation of
momentum balance must be considered. In so doing there results a new elastic
force that
does no work\cite{magnetic} and that consequently cannot be obtained
on the basis of energy considerations alone. For a crack moving in a thin
elastic plate, consideration of three dimensionality makes this force operative
and provides a scenario, some of it qualitative, some of it quantitative, that
appears to explain many of the observed experimental facts.

The equations of dynamic elasticity in a medium of dimensionality $d$ and
density $\rho$ can be obtained as extrema of the action
functional
\beq
S = \frac 12 \int dt d^d x (\rho {\dot u}^2 - \sigma_{ij} s_{ij} )
\eeq
where $\sigma_{ij}$ is the stress, $s_{ij}$ the strain, $i,j=1,\dots,d$ and
$\vec u (\vec x,t)$ the particle deviation at time $t$ from an equilibrium
position $\vec x$. When stress and strain are
linearly related, $\sigma_{ij} = c_{ijkl} s_{kl}$, the well known linear
equations can be written in conservation-law form\cite{oldeshel,lund88}:
\bea
\label{eq:enmom}
\pder{T^{00}}{t} + \pder{T^{0i}}{x^i} & = & 0 \\
\pder{T^{j0}}{t} + \pder{T^{ji}}{x^j} & = & 0 ,
\eea
with
\bea
\label{eq:comp}
T^{00} & = & \frac{\rho}{2} \left( \pder{u}{t} \right)^2 + \frac 12
c_{ijkl} \pder{u^i}{x^j} \pder{u^k}{x^l} \\
T^{0j} & = & -c_{jikl} \pder{u^i}{t} \pder{u^k}{x^l} \\
T^{j0} & = & -\rho \pder{u^i}{t} \pder{u^i}{x^j} \\
T^{ji} & = & c_{klmi} \pder{u^k}{x^l} \pder{u^m}{x^j}  \nonumber \\
 & & \qquad + \frac 12 \left(
\rho \left( \pder{u}{t} \right)^2 - c_{klmn} \pder{u^k}{x^l}
\pder{u^m}{x^n} \right) \delta_{ji}
\eea
The components of the energy momentum tensor $T$ have the following
interpretation: $T^{00}$ is the energy density, $T^{0j}$ is the energy
flux, $T^{j0}$ is the momentum density and $T^{ji}$ is the momentum flux.

If Eqns. (\ref{eq:enmom}) are integrated over a volume $\cal V$ that is
bounded by a surface $\cal S$ that is possibly moving with velocity
$\vec V$ one obtains energy conservation
\beq
\label{eq:encons}
\frac{d}{dt} \int_{\cal V} d^dx T^{00} = \int_{\cal S} dS_j T^{0j} -
\int_{\cal S} dS_j V^j T^{00}
\eeq
and (linear) momentum conservation
\beq
\label{eq:momcons}
\frac{d}{dt} \int_{\cal V} d^dx T^{i0} = \int_{\cal S} dS_j T^{ij} -
\int_{\cal S} dS_j V^j T^{i0}  .
\eeq

If Eqn. (\ref{eq:encons}) is applied to the case of a crack advancing in
a straight line, we get the usual equation of motion for the crack
tip\cite{freund}. There is energy flow into the tip, which is equated to
the surface energy that is created as the crack advances. The volume
$\cal V$ is the volume of the whole sample, and the surface $\cal S$ is
composed of a small (i.e. smaller than any relevant length such as
sample thickness or radiation wavelength) cylinder surrounding the crack
tip, the lips of the crack, and the boundaries of the sample. The
contribution of the small cylinder around the tip is finite because
stresses behave like $1/\sqrt{r}$, with $r$ distance to the tip, near
the crack. The crack lips do not contribute because they are at rest
and, by definition, they are traction free. Remote surfaces do not
contribute as long as radiation reaction is neglected.

It is a simple matter to check that, for a crack advancing along the
$x^1$ direction at velocity $V$ the following is true:
\beq
\label{eq:onedim}
V\frac{dP^1}{dt} =  \frac{dE}{dt}
\eeq
where
\bea
E & = & \int_{\cal V} d^dx T^{00} \\
P^i & = & \int_{\cal V} d^dx T^{i0}
\label{eq:defs}
\eea
so that the momentum equation and the energy equation provide the same
information. Now, if the crack tip motion is not constrained to a straight
line,
the momentum equations provide the additional information needed to determine
the motion. In two dimensions, one equation would not suffice and less so in
three.

Consider now the normal modes of vibration of an elastic plate
in three dimensions\cite{achenbach}. For a thin plate of thickness $h$ there
are, to
leading order, one dilatational mode with frequencies $n\pi \alpha /h$ and two
shear modes with frequencies $n \pi \beta /h$, where $\alpha$ is the speed of
sound and $\beta$ the speed of shear waves, both in bulk, and $n=1,2,3,\dots$.
(Figure 1). When a crack propagates along the $x^1$ direction, say, because of
static loading along $x^3$, the modes
involving motions parallel to the crack faces will be unaffected. There is,
however, one shear mode that involves motion along the $x^3$ direction that
will
be incompatible with the traction-free boundary condition that defines a
crack. Consequently this mode will be modified by the presence of the crack
and stress concentration, in addition to the one already present due to the
static loading, will result. The question naturally arises as to how this new
stress concentration at the crack tip will affect its dynamics. Of course,
these
 vibrating modes do not exist in two dimensions.

In order to give a quantitative assesment of this effect we consider the crack
surface near the tip as a continuous distribution of infinitesimally
small dislocations. These dislocations are defects in a continuum, and they are
not related to a (possibly non-existent) underlying crystal structure. That
this
makes sense is well-known in the static case\cite{cracdisloc} and has also
been used in dynamics, to show that a crack may, under certain circumstances,
acquire inertia\cite{lunddiaz}. The basic idea is that the infinitesimal
dislocations arrange themselves so as to negate the externally applied stress
and enforce the no traction boundary condition at the crack surface. The exact
functional behaviour of the crack opening displacement that will result in the
case at hand is not needed
at the current stage of development of the theory, although it presents itself
as an attractive challenge for future research. We only need to know that {\it
some} such distribution exists.

Consider then an edge dislocation of (infinitesimal) Burgers vector $\vec b$
moving with velocity $\vec V$ in an elastic medium whose (small) particle
displacements away from an equilibrium position are $\vec U (\vec x, t)$. The
relative orientation of $\vec b$ and $\vec V$ is immaterial. The
time dependence of $\vec U$ does {\it not} include a possible motion of the
body as a whole, since that would not be a small deviation away from an
equilibrium position. The dislocation generates around it a well known\cite{hl}
particle displacement $\vec u (\vec x , t)$ in which strain and particle
velocity are related, at short distances, by ($a,c = 1,3$)
\beq
 \pder{u_a}{t} = V_c \pder{u_a}{x^c}
\eeq
The point is to compute the force
\beq
\label{eq:force}
F^j \equiv \frac{d P^j}{dt}
\eeq
given by (\ref{eq:momcons}) and (\ref{eq:defs}), for which the relevant
integrals have to be evaluated around a small circle enclosing the
dislocation. Assuming the medium to be linear down to this small circle, the
particle displacements will be the sum
\beq
\vec u + \vec U
\eeq
of dislocation generated plus externally imposed ones. Since the
components of the energy momentum tensor $T$ are bilinear functions of
derivatives of displacements, they will decompose naturally in three terms: one
involving only $\vec u$, one involving only $\vec U$, and one involving both
$\vec u$ and $\vec U$  which is the only one of interest:
\beq
\label{eq:mix}
T^{j0} = -\rho \left( \pder{u^i}{t} \pder{U^i}{x^j} + \pder{U^i}{t}
\pder{u^i}{x^j} \right)
\eeq
and similarly for $T^{ij}$. Substitution of the field $\vec u$ for an edge
dislocation\cite{hl} into (\ref{eq:momcons}), (\ref{eq:defs}) and
(\ref{eq:force}) using (\ref{eq:mix}) gives the result
\beq
F^j = F_{PK}^j + F_V^j
\eeq
where
\beq
F_{PK}^j = b_i \Sigma_{ij} \epsilon_{jkm} {\hat t}_m
\eeq
with $\hat t$ the unit vector along the dislocation line and $\Sigma_{ij} =
c_{ijkm} \partial U_k /\partial x_m $ the external stress, is the usual
Peach-Koehler force. In addition there is a force\cite{veldepforce}
\beq
F_V^j = \rho \epsilon_{jkm} V_k {\hat t}_m b_i {\dot U}_i
\eeq
that acts only if the dislocation is moving ($\vec V \ne 0$), if the elastic
medium itself is also independently moving ($\dot{\vec U} \ne 0$), and is
perpendicular to the direction of motion and consequently does
no work. For a Burgers vector corresponding to a mode-I crack, the static
loading gives a Peach-Koehler force along the cleavage plane. When summed over
the many infinitesimal dislocations that make up the crack tip, it gives the
usual work-performing force that is considered in the standard
treatement  of crack-tip dynamics .
It does not have a component perpendicular to the motion. The vibrational
modes of the plate that have been previously considered give a vanishing
Peach-Koehler force. The force ${\vec F}_V$, on the other hand, is nonvanishing
only for the shear mode that was already argued to be concentrated at the tip
in order to satisfy the proper boundary conditions along the crack faces. The
static loading of course does not contribute to it.

As it was mentioned at the beginning of this communication, recent
experiments \cite{p1,p2,p3,p4} indicate that cracks in glass and plexiglass
suffer an instability at a critical velocity, beyond which there are strong
 sound emissions, strong velocity fluctuations, and strong surface
 heterogeneities. The limiting velocity is well below the theoretically
expected
 Rayleigh wave velocity.
We are now in a position to propose the following scenario to explain those
observations:
\begin{enumerate}
\item
The crack starts propagating, and right away it starts emitting elastic
(both longitudinal and shear) radiation. The pull tabs possibly also
radiate at the beginning. The crack accelerates.
\item
Normal modes (in thickness) of vibration of the plate are established,
and they are responsible for a new oscillating force that does
no work on the crack and is proportional to the tip's velocity. The fundamental
mode will dominate but at sufficiently
high crack velocity the forcing may be enough to also excite higher modes. To
leading order they are the modes of an infinitely extended slab. Next order
corrections should take into account the finite size and shape of the sample.
\item
The crack responds by oscillating to the tune of this new periodic (possibly,
multiply periodic)
forcing and, hence, emitting radiation (both shear and longitudinal) primarily
at the same frequencies it is being forced. This radiation reinforces the
normal
modes that, in turn, feed back on the tip. This is the instability.
\item
In a non-amorphous material such as a crystal, the surface energy may be
higher for out of plane crack motions and thus this mechanism may be non
operative.
\item
Terminal speed (i.e. average speed along line of propagation) is determined
by energy spent by external stress. Since the crack path is effectively
increased, the average speed of advance is decreased.
\end{enumerate}
The above scenario is universal in the sense that it applies to any brittle
material that is described by linear elasticity. At this stage, there is no
calculation of a transition velocity, but a reasoning that indicates why there
should be an instability as the crack velocity increases, There is, however, a
clear consequence that sound radiation, velocity oscillations and surface
roughness are different manifestations of the same phenomenon, whose spectra
should be peaked at the normal modes of the elastic plate. These consequences
should be amenable to experimental
verification, for example by studying the effect of sample thickness and of
sample ensonification.

To conclude: it has been established that elastic singularities are subject to
velocity dependent forces that do no work. On this basis, a scenario has been
proposed to explain recent observations of dynamic crack instabilities. The
proposed scenario includes the three dimensionality of the problem as an
essential ingredient. It provides a qualitative physical origin for the
instability,
and a quantitative prediction for the frequencies that should be notorious in
the resulting spectra of sound, surface and velocity fluctuations, namely, the
eigenfreuencies of the thin plate that is being fractured.

Stimulating discussions with S. Ciliberto are gratefully acknowledged. Part of
this research was done while on a visit to the Centre Emile Borel, wherein a
fruitful conversation with M. Benamar was held. This work was supported by
Fondecyt Grant 1940429, CCE Contract CI1*CT91-0947 and the Andes Foundation.

\section*{Caption for figure}
 \paragraph{Figure 1} A thin elastic plate of thickness $h$ has, to leading
approximation,
three normal modes of elastic oscillation: one (a) is dilatational, with
eigenfrequencies $n\pi \alpha /h$ and two (b-c) are shear, with
eigenfrequencies
$n\pi \beta /h$, where $\alpha$ is the speed of sound, $\beta$ is the speed of
shear waves, and $n= 1,2,\dots$ When a crack propagates along the
$x^1$direction, for instance because of static loading along $x^3$, the shear
mode indicated in (c) is incompatible with traction-free boundary conditions
along the crack faces, leading to a new stress concentration at the tip that
will modify the crack dynamics.


\begin{thebibliography}{99}
\bibitem{magnetic} In a different physical context, an example of such a force
is that exerted by a magnetic field on a moving electric charge.
\end{thebibliography}
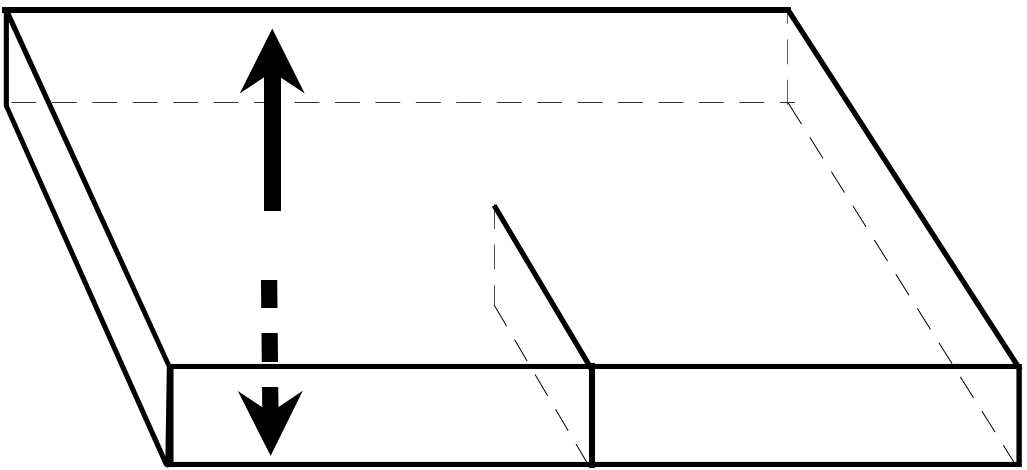
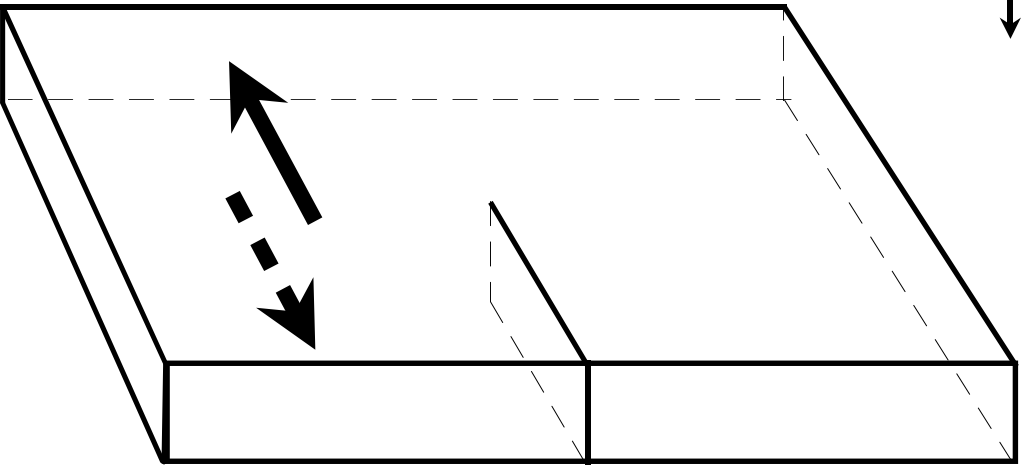
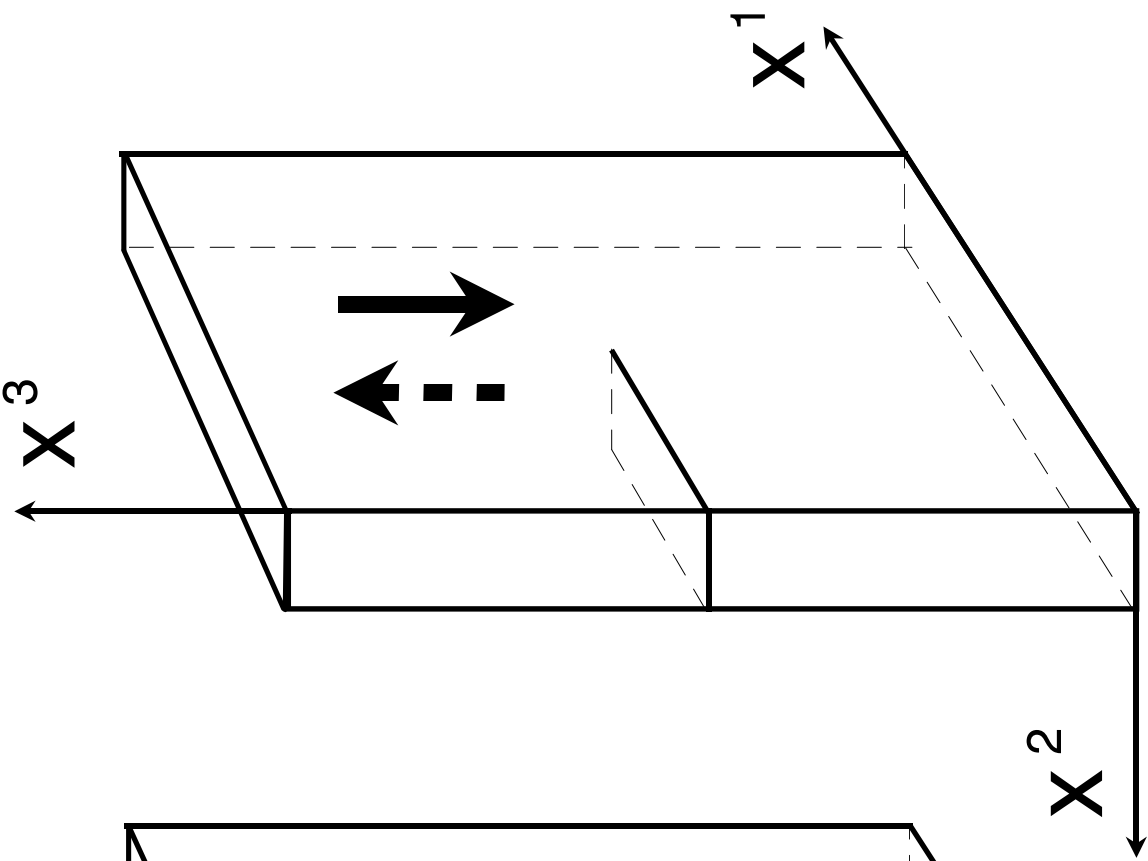
\end{document}